# Experimentally-validated multi-slice simulation of electron diffraction patterns

Xinke Xiao[a], Tianle Ma[a], Lingxuan Shao[b], Jun Liu[c], Qiwei Shi[b*], Canying Cai[a*], Stéphane Roux[d]

[a] School of Materials Science and Engineering, Xiangtan University, Xiangtan 411105, China

[b] SJTU-ParisTech Elite Institute of Technology, Shanghai Jiao Tong University, Shanghai 200240, China

[c] School of Materials Science and Engineering, Shanghai Jiao Tong University, Shanghai 200240, China

[d] Université Paris-Saclay, ENS Paris-Saclay, CNRS, LMPS - Laboratoire de Mécanique Paris-Saclay, F-91190 Gif-sur-Yvette, France

Emails: sqw@sjtu.edu.cn, cycai@xtu.edu.cn

## Abstract

High-Resolution Electron Backscatter Diffraction (HR-EBSD) has advanced rapidly in recent years, significantly improving elastic strain measurements and dislocation density evaluation with submicron spatial resolution. To achieve better accuracy in the measurements, high-quality dynamical simulation patterns are required to be matched with experimental ones. Currently, the most widely used pattern simulation method, the Bloch Wave method (BW), can accurately predict the positions and brightness of Kikuchi poles and bands, but is intrinsically limited to perfect crystal structures. Another simulation scheme, the multi-slice method (MS), follows the evolution of electron waves as they travel through the sample. MS is advantageous in simulating various defect structures with more diffraction details. Yet, it is mainly considered for theoretical developments and has not been compared to experimental data.

This paper optimizes the MS method by abandoning the high-energy hypothesis and utilizing higher-order Taylor expansions to approach the forward-only Schrodinger equation. Experimental EBSD patterns of polycrystal Al-Mg alloys are used to challenge MS simulations as a reference for indexation. It is demonstrated that the 5th-order expansion of MS, referred to as MS5, achieves a good balance between computational cost and pattern precision. A tailored isotropic distortion correction model and standard stereographic triangle reconstruction enhance the precision of MS5 to be comparable with BW. To the best of our knowledge, this study provides the first comparison of MS EBSD simulations with experimental data. It opens new possibilities for EBSD characterization, such as reproducing diffraction patterns of crystals with various defects.

**Keywords**: EBSD, Multi-Slice Method, Pattern Indexation, Digital Image Correlation

# 1. Introduction

Electron Backscatter Diffraction (EBSD) has undergone rapid development over the past few decades and is now widely adopted as an analytical technique for materials characterization [1, 2]. It is widely used for crystal phase identification [3, 4], orientation determination [5, 6], and stress-strain analysis [7, 8]. Automated EBSD indexing methods rely on image analysis algorithms. The Hough Transform has been a popular method since the early work of Lassen [9], and is still widely used for EBSD pattern indexing. However, specific patterns remain challenging to index, such as those of severely plastically strained materials with hardly recognizable diffraction bands, or those observed at grain boundaries when two or more diffraction patterns overlap. Simulated EBSD patterns are key for analyzing these challenging cases.

In recent years, the HR-EBSD technique proposed by Wilkinson et al. has gained popularity for its ability to retrieve elastic strains via accurate cross-correlation of experimental diffraction patterns [10]. S. Villert et al. applied the cross-correlation algorithm to analyze experimental results from four-point bending of single-crystal silicon and a Si-Si$_{1-x}$Ge$_x$ system, comparing them with finite element simulation results to evaluate the accuracy and reliability of the algorithm [11]. However, both experiments rely on the presence of strain-free regions within the sample — which may be difficult to find in practice — to obtain full-field stress values. The Electron Backscatter Pattern (EBSP) from a near-undistorted region within the same sample is typically used as the reference pattern. The comparison then yields relative elastic strain and rotation data. Resorting to a simulated strain-free reference pattern is an appealing yet challenging approach to obtaining high-precision *absolute* strain data.

Electron diffraction theory is typically divided into two approaches: kinematic and dynamic. Kinematic simulations predict precisely the location of Kikuchi bands. Still, because they neglect multiple scattering, the predicted intensity distribution is unreliable and insufficient for absolute stress measurements. In contrast, dynamic simulations (*i.e.,* including multiple scattering) offer a more faithful description that better matches experimental patterns and are therefore a superior choice for HR-EBSD [12]. Winkelmann et. al. proposed a refinement of the electron/crystal lattice interaction model [13], leading to an even better match with experimental Kikuchi patterns. Consequently, the precision of quantitative analysis was enhanced [14, 15]. A 'master pattern' summarizing the diffraction details of all possible crystal orientations can be dynamically simulated, which enables the use of Digital Image Correlation (DIC) to determine the projection parameters of the experimental pattern. Consequently, a different indexing approach – the dictionary method – has been developed [16]. This method can provide reliable indexing results even in cases of poor EBSD pattern quality.

The standard methods for calculating dynamical electron diffraction patterns are the Bloch Wave method (BW) and the Multi-Slice Method (MS). The BW method is directly derived from the Schrödinger equation [2] and can produce high-quality diffraction patterns under the assumption of periodicity. In recent years, Winkelmann used the BW method to propose a multi-beam dynamical simulation capable of reproducing the intensity distribution in EBSD patterns [17]. Furthermore, Winkelmann introduced anisotropy for backscattered electrons through optimizing diffraction theory formulae and reproduced the Excess-Deficiency (E-D) effect using a phenomenological model [18, 19]. Zhu and De Graef developed an approximate model for the inclusion of dislocations in the simulation of a Kikuchi pattern [30] by projecting the master pattern from periodic crystal models. Hence, the BW method is intrinsically limited to simulating perfect crystals and cannot directly calculate EBSD patterns with defects.

In contrast, the MS method can calculate imperfect crystal structures and provide more diffraction details than the BW method. The Conventional MS method (CMS) was initially proposed by Cowley and Moodie [20]. Later, Van Dyck demonstrated that the CMS method could be directly derived from the Schrödinger equation [21] and introduced a new multislice theory – the Real Space method (RS) [22]. Cai et al. compared CMS and RS methods, finding that RS held advantages in both speed and precision [23]. Subsequently, Chen et al. established a precise multislice theory [24, 25], and Cai et al. successfully obtained a numerical solution, confirming that the Revised Real Space method (RRS) has higher precision [26]. Liu compared the details between the RS and RRS methods and qualitatively analyzed the RRS and third-order MS method, concluding that their differences were negligible for experimental indexing [27]. Subsequently, Cai's group used the RRS method to simulate the effects of edge/screw dislocations and twin structures on EBSPs, respectively [28, 29]. However, the MS method has not been used to analyze experimental patterns quantitatively or to index them.

The present study aims to optimize the MS method to a sufficient level of precision, thereby opening new possibilities for EBSD characterization. Section 2 explains the MS method with different precision levels. Qualitative and quantitative analyses are conducted in Section 3 to compare the simulated patterns obtained using MS, BW, and kinematic methods with the experimental patterns. Residual fields offer a very discriminative tool for this purpose. Section 4 reports the first crystal orientation indexation based on MS simulations and evaluates its precision levels. This research guides the subsequent comparison of MS simulations with experimental EBSPs and also opens new perspectives, such as quantifying defect structures using HR-EBSD.

## 2. Theory and Computational Models

The formation of EBSPs is a complex process involving coherent, elastic, and inelastic scattering of electrons. The MS method for EBSP simulation divides the sample into numerous thin slices, as shown in Fig. 1. As the initial incident wave moves forward, the exit wave function of the preceding slice serves as the incident wave function for the subsequent slice. The interaction between electrons and the sample is calculated slice by slice. By summing up the diffraction patterns generated from each slice, the final EBSD diffraction pattern is obtained. The details are provided below.

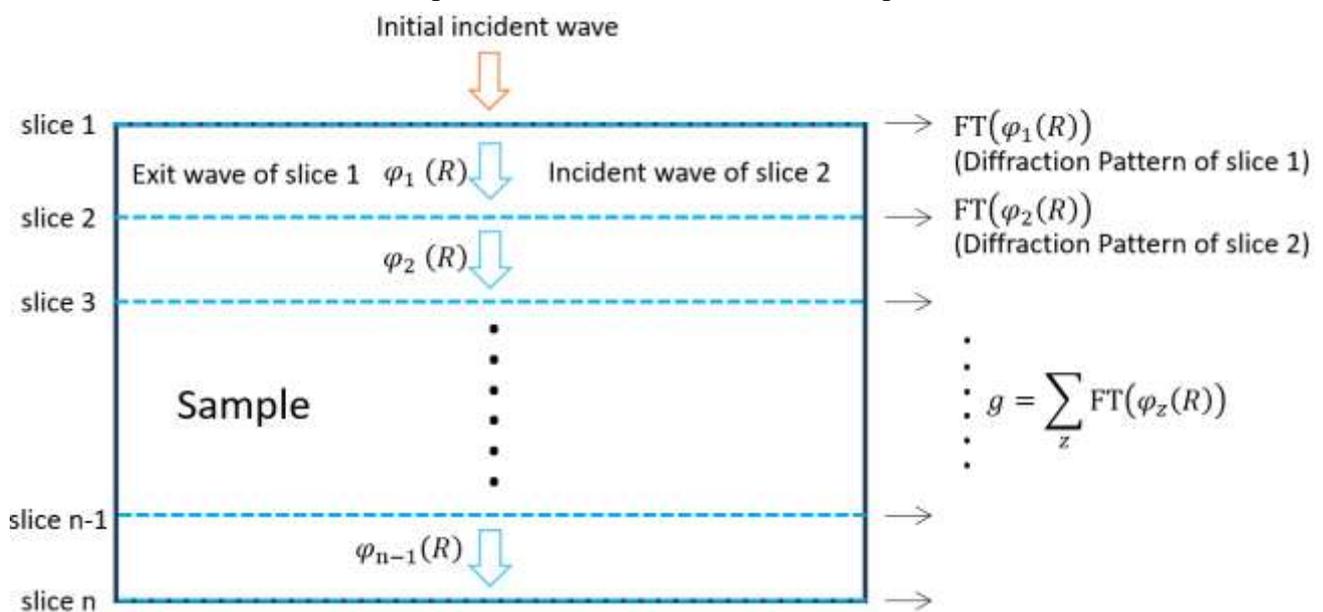

**Fig. 1.** Schematic diagram of the MS method.

MS focuses on the forward propagation of electron waves, while neglecting all backscattering ones [24], a scenario more similar to transmission Kikuchi diffraction (TKD). Only the electrons carrying crystallographic information are considered as they exit the sample to form the final EBSD pattern. The fundamental theory of the MS method is derived from the Schrödinger equation, yielding Equation (1) as detailed by Cai [26].

$$\frac{\partial \varphi(R,z)}{\partial z} = 2\pi i k_0 \left[ \sqrt{1 + \frac{1}{(2\pi k_0)^2}\Delta + \frac{\sigma}{\pi k_0}U(R,z)} - 1 \right] \varphi(R,z) = \bar{H}(R,z)\varphi(R,z) \quad (1)$$

$\varphi(R,z)$ is the slice wave function in the plane $R(x,y)$ at depth $z$, $k_0$ the wave vector of the incident wave, $\Delta = (\partial^2/\partial x^2) + (\partial^2/\partial y^2)$ the two-dimensional Laplacian operator, $\sigma$ the interaction constant defined as $\sigma = \frac{2\pi e \lambda m}{h^2}$, and $U(R,z)$ the crystal potential. Note that the scattering factor is taken as the square of the atom weight in the BW method [30], while in MS simulations, the electron scattering is calculated based on $U(R,z)$ by Equation (1). The numerical solution of Equation (1) is obtained by applying a Taylor expansion of $\bar{H}(R,z)$. The Taylor expansion is derived up to the 5th order, and the general expression is given in Equation (2)[1]. As the order of the Taylor expansion increases, the results from the MS method become progressively more accurate numerically. The 1st- to 5th-order MS methods refer to truncating the Taylor expansion at the corresponding order and are denoted MS$n$, with $n$=1,…, 5 corresponding to the truncation order. Note that the previously mentioned MS methods for EBSD simulation belong to this MS series. For example, the real space (RS) method proposed in Ref. [23] corresponds to MS1, while the revised real space (RRS) discussed in Ref. [27] is exactly MS2. The current work proposes a generic MS$n$ method that could be truncated at a user-set order $n$,

$$\bar{H}(R,z) = \left\{ i\left[\frac{\Delta}{4\pi k_0} + \sigma U(R,z)\right] - \frac{i}{4\pi k_0}\left[\frac{\Delta}{4\pi k_0} + \sigma U(R,z)\right]^2 + \right.$$

$$\frac{i}{8(\pi k_0)^2}\left[\frac{\Delta}{4\pi k_0} + \sigma U(R,z)\right]^3 - \frac{5i}{64(\pi k_0)^3}\left[\frac{\Delta}{4\pi k_0} + \sigma U(R,z)\right]^4 \quad (2)$$

$$\left. + \frac{7i}{128(\pi k_0)^4}\left[\frac{\Delta}{4\pi k_0} + \sigma U(R,z)\right]^5 + \cdots + \frac{(-1)^{n-1}(2n-3)!! \, i}{(n!)\left(2^{n-1}\right)(\pi k_0)^{n-1}}\left[\frac{\Delta}{4\pi k_0} + \sigma U(R,z)\right]^n \right\}$$

The MS model does not involve a high-energy approximation. The chosen convergence criterion used at each slice was that the relative correction value of the iterated wave function dropped below $10^{-7}$. EBSP is formed by a collection of backscattered electrons diffracted from different sample depths. Hence, the pattern of each slice, calculated by the Fourier transform of $\varphi(R,z)$, is summed up to estimate $g$, the final EBSP defined on an approximate Ewald sphere in the reciprocal space,

$$g = \sum_z \mathrm{FT}(\varphi(R,z)) \tag{3}$$

where FT stands for the Fourier Transform. As the Taylor expansion order increases, $g$ increasingly approaches a perfect Ewald sphere. The Kikuchi bands on $g$ exhibit curvature, and their projection from the sphere onto a (detector) plane is necessary for comparison with the experimental pattern.

---

[1] This corrects a previous coefficient error in the 4th-order term of the Taylor expansion, changing it from 1/16 to 5/64 [26].

As shown in Fig. 2, let $P$ be a point on a sphere of radius $r$. The intersection of the line connecting $P$ and the north pole N with the equatorial plane yields the new coordinate $Q$, which is obtained via polar stereographic projection of the spherical coordinate $P$. Equation (4) describes the correspondence between the two coordinate systems.

$$(x, y, |z|) = \left( \frac{2u \cdot r^2}{u^2 + v^2 + r^2}, \frac{2v \cdot r^2}{u^2 + v^2 + r^2}, \sqrt{r^2 - x^2 - y^2} \right) \tag{4}$$

Through this correspondence, the pattern on the Ewald sphere can be projected onto the equatorial plane using two-dimensional interpolation. Subsequently, the Integrated Digital Image Correlation (IDIC) algorithm [31] is used to project the patterns from the equatorial plane onto the screen plane, simultaneously calibrating the projection parameters, including the Euler angle triplets $(\varphi_1, \phi, \varphi_2)$ and the projection center coordinates $(x^*, y^*, z^*)$. This projected pattern can be quantitatively compared with the experimental pattern.

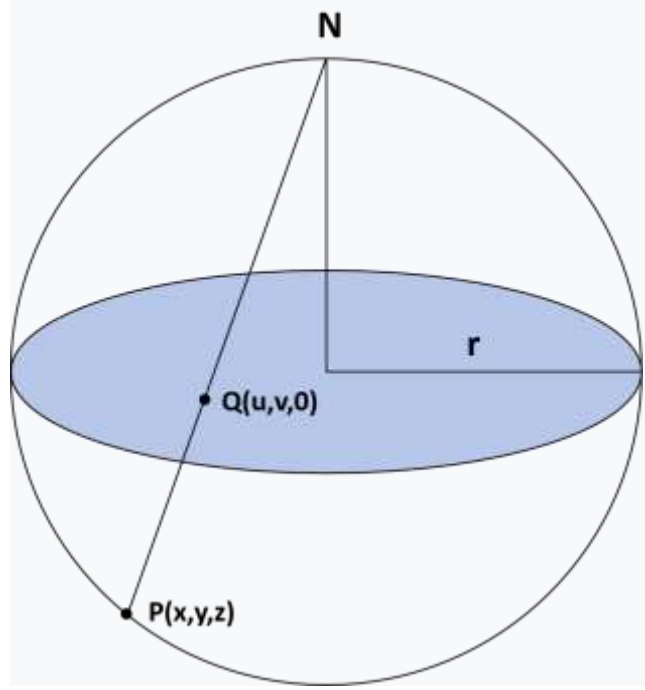

**Fig. 2.** Schematic diagram of spherical coordinate transformation to polar stereographic projection coordinates.

To calculate and compare the differences between patterns generated by MS methods at different orders, the Al structure is used, with a lattice constant of 0.405 nm. By introducing the elastic scattering parameter of Al [32], we calculate the crystal structure factor $F(\vec{u})$. Subsequently, the crystal potential $U(\vec{r})$ can be calculated using equation (5):

$$U(\vec{r}) = \frac{1}{V_c} \int_{u=-\infty}^{\infty} F(\vec{u}) \exp(-2\pi i (\vec{u} \cdot \vec{r})) \, d\vec{u} \tag{5}$$

where $V_c$ is the volume of the unit cell in real space, $\vec{r}$ the real-space vector, $\vec{u}$ the reciprocal vector in reciprocal space, and $u = |\vec{u}|$. The incident electron beam energy was set to 20 kV, and its incident direction was [001]. A 6-fold cell expansion was performed in the directions perpendicular to the incident beam, specifically along the [010] and [100] directions. In these two directions, each unit cell is discretized into $n_x = n_y = 192$ bins in both directions. The definition of the final EBSD simulated pattern is set to 1152 × 1152, the same as the number of expanded bins (6×192).

For the [001] direction parallel to the incident beam, a 40-fold cell expansion was performed, resulting in a total model thickness of 16.2 nm after expansion. The number of slices per unit cell along this direction can be calculated using the optimal slice thickness formula proposed recently in [33], namely

$$\varepsilon \leq \varepsilon_0 = 4k_0 \Delta x^2 C \tag{6}$$

where $\varepsilon_0$ is the optimal slice thickness that minimizes computation time, $\Delta x$ the sampling interval (which is the slice thickness perpendicular to the incident beam), $C$ a real number that depends on the nature of the numerical algorithm (set to 1 here). The slice thickness has a significant impact on computation time. Adopting excessively thick slices may lead to convergence problems. The optimal number of slices per unit cell is 1954, as Equation (6) predicted. In the front view of the computational model, as shown in Fig. 3b, the intensity distribution of the initial delta wave function is illustrated by the black line, i.e., a focused electron beam with the same expanded bin definition (1152×1152). Corresponding to the electron beam conical spread. Periodic boundary conditions are applied to the propagation of electron waves.

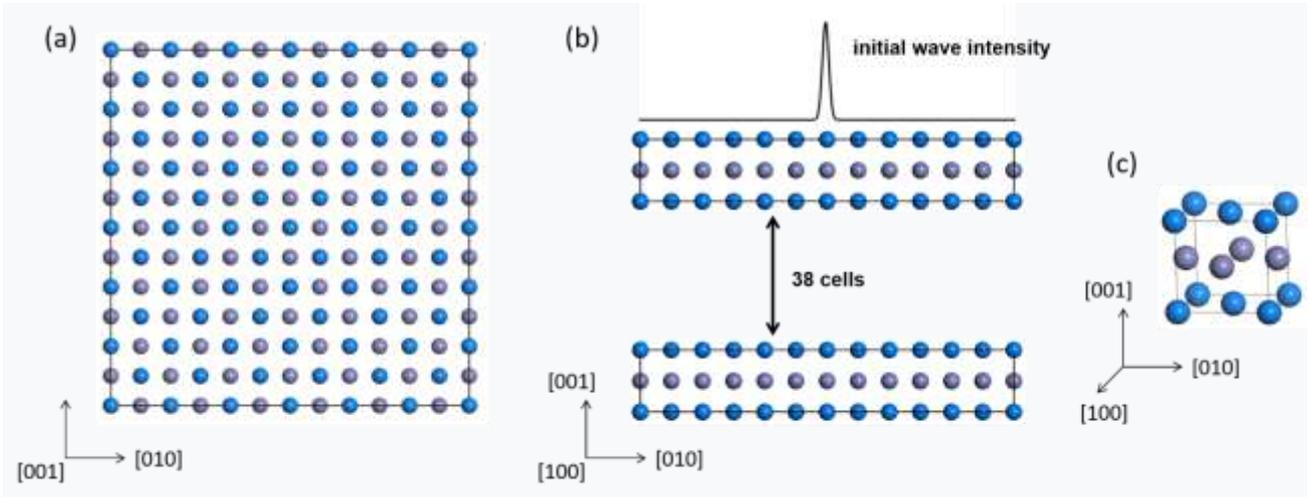

**Fig. 3.** Model for calculating the EBSD pattern of Al in the [001] direction, showing the top view (a) and the front view (b). (c) a unit FCC cell in a three-dimensional view of the simulation model. Atoms of different layers of the unit cell are shown in different colors.

The MS simulations are programmed in Matlab. They are run on a computer equipped with an Intel Xeon Gold 6130 CPU and an NVIDIA Quadro P4000 GPU. Vectorization and GPU acceleration were applied in Matlab to accelerate the calculation. Table 1 lists the computation time required for MS methods at different orders. For comprehensive reasons, the computation time of the MS5 master pattern (to be defined in Section 3.2) in Matlab and the BW master pattern in EMsoft, both at the same resolution, are also compared. The additional calculation required by the MS5 master pattern only incurred negligible extra time.

**Table 1:** Computation time (in hours) required for different orders of the MS Method, MS5 master pattern by Matlab, and BW master pattern by EMsoft.

|  | MS1 (RS) | MS2 (RRS) | MS3 | MS4 | MS5 | MS5 master pattern (MS5 + radial distortion correction + symmetry operation) | BW |
|---|---|---|---|---|---|---|---|
| 128 slices | 0.87 | 1.45 | 2.12 | 3.08 | 3.66 | - | - |
| 192 slices | 3.69 | 7.16 | 13.24 | 33.00 | 101.39 | 101.39 | 2.01 |

The Matlab code took only 0.51 s to calculate the crystal potential. Nevertheless, for small computational models such as the 4-atom Al unit cell, most of the computation time was spent calculating the interaction between the electron wave function and the crystal potential within each slice. Comparing the calculations with $n_x = n_y = 128$ per unit cell, the computation time was essentially linear with the expansion order. However, with $n_x = n_y = 192$ bins, the computation time increased significantly and no longer followed a linear relationship with the Taylor expansion order. For the MS1-3 methods, computation time roughly doubled with each increase in order; for the MS3-5 methods, MS method computation time increased by a factor of three with each subsequent order. This rapid time-rate inflation is mainly due to the increasing number of iterations required to converge for each slice. Simultaneously, the increased number of slices in the [001] direction, calculated by Equation (6), leads to more calculation steps. Using finer resolution in the [100] and [010] directions in real space enlarges the field of view of the simulated pattern in reciprocal space, as shown in Fig. 4. This is because distances in real and reciprocal spaces are inversely proportional. The lines of different colors in Fig. 4 indicate the correct position of the Kikuchi band center lines of different crystal plane families in the spherical projection. The EBSD pattern simulated with 192 bins matches these lines at the periphery better. For reasons of calculation time and view field, all the following MS simulations use 192 bins.

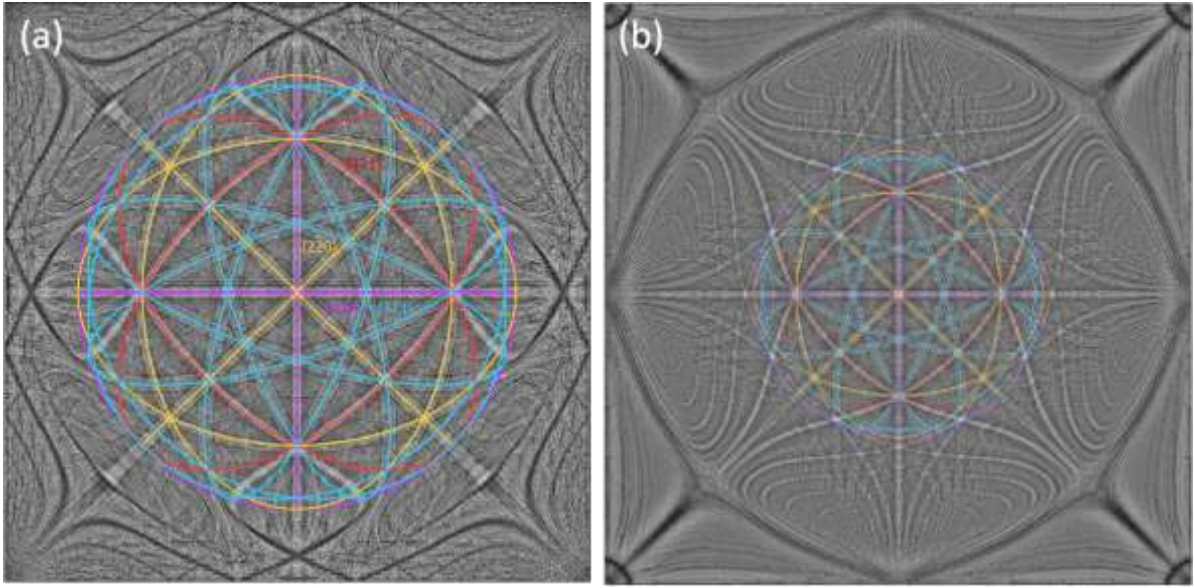

**Fig. 4.** Comparison of EBSD pattern simulated with 128 bins (a) and 192 bins (b) in the *x-y* plane.

## 3. Validation of MS simulations
### 3.1   Raw MS simulations

Fig. 5 illustrates the diffraction patterns of Al, calculated using the MS$n$ methods with $n = 1, \ldots, 5$. All MS simulated patterns show a highly contrasted center and a fading peripheral. This is because the initial wave function used in the MS calculation is a focused electron beam with high central intensity that gradually decreases outward. Consequently, the portion of the beam closer to the incident center interacts more with the crystal, leading to higher pattern intensity in that region. The theoretical positions of the major K-bands are shown as colored lines on the left side of Fig. 5 to facilitate visual comparisons.

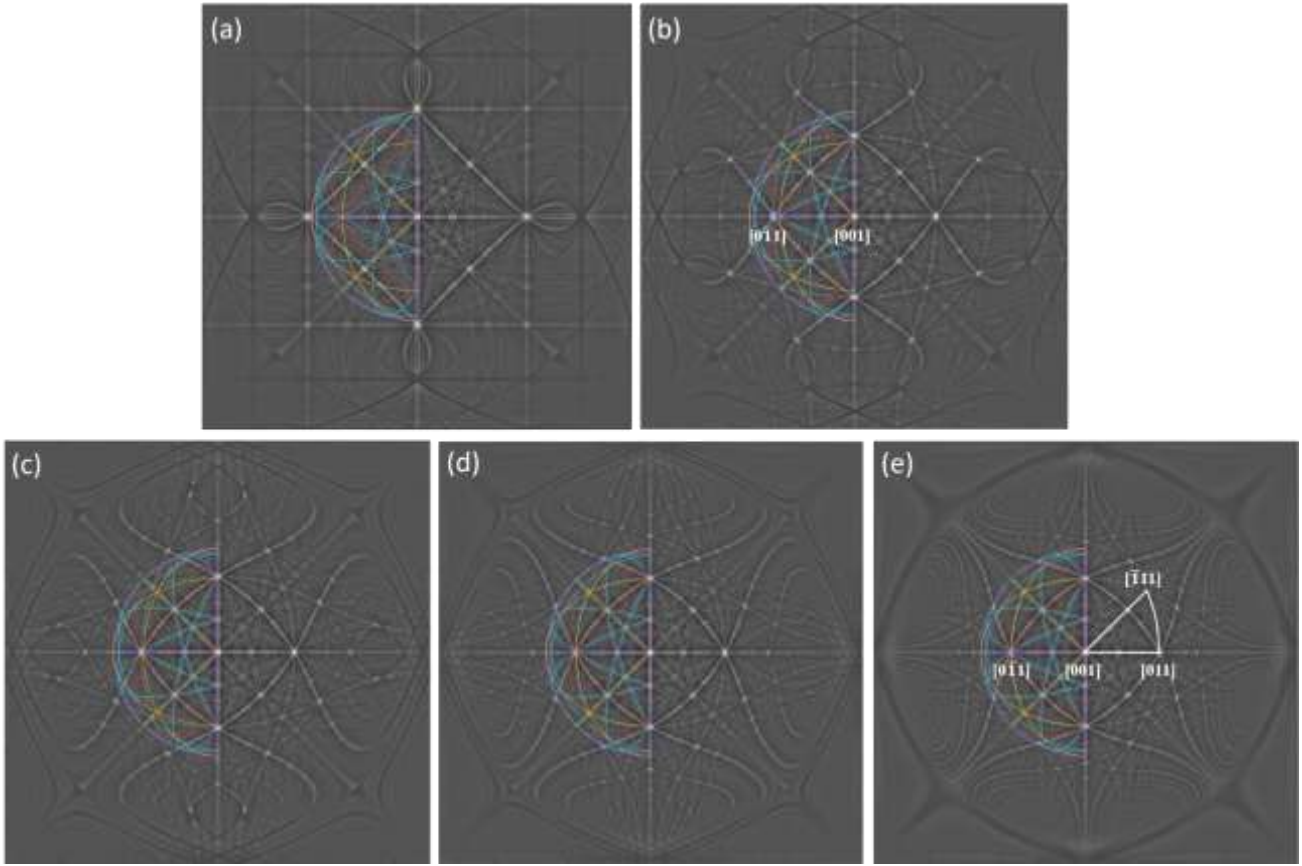

**Fig. 5.** (a)-(e) Diffraction patterns of Al in the [001] direction calculated using the MS method of orders 1 to 5.

A significant difference exists between the MS1 and MS2 results. However, comparing the more similar MS2-5 patterns reveals a trend of contraction towards the center. As the MS method order increases, more peripheral bands become visible within the same field of view, while the central region shows relatively minor changes. This is demonstrated by the [0$\bar{1}$1] Kikuchi pole in Fig. 5b and Fig. 5e. As the order is increased from 2nd to 5th, the [0$\bar{1}$1] Kikuchi pole exhibits a slight, almost indiscernible displacement toward the center. Fig. 5e clearly shows the [0$\bar{1}$1] Kikuchi pole moving a small distance toward the center. This phenomenon suggests that a higher-order expansion is needed to locate the peripheral bands correctly.

As the numerical solution of the Taylor expansion becomes increasingly precise for higher-order expansions, the MS method simulation pattern increasingly approaches a perfect spherical projection. In the MS4-5 patterns, shown in Fig. 5d-e, a circular boundary exists at the periphery, where many nearly parallel Kikuchi bands form the edge. The peripheral Kikuchi bands in the MS4-5 patterns are essentially 'compressed', creating a circular edge. Since the MS method only considers pure forward scattering of electrons during the calculation and neglects changes in electron energy, no diffracted electrons are expected to fall outside the Ewald sphere. Thus, the EBSD pattern disappears beyond this circular boundary, and the simulation results are consistent with expectations.

To more intuitively visualize the influence of the different orders of MS on EBSD simulations, we subtracted the MS5 pattern from the MS1-4 ones, yielding the intensity difference maps shown in Fig. 6. The difference map between the MS1 and MS5 patterns, shown in Fig. 6a, exhibits a near-zero difference only in a small central Kikuchi pole region. In contrast, the pattern shows a significant residual away from the center. As the order increases, the central area with near-zero difference progressively expands. In the difference map between the MS4 and MS5 patterns, shown in Fig. 6d, the region within the black dashed circle shows zero difference. This part of the pattern remains constant as the MS method order increases, indicating that the EBSD pattern within this region is reliably calculated. The regions enclosed by the $[001]$, $[011]$, and $[\bar{1}11]$ Kikuchi poles (marked by solid-lines) in Fig. 5e and Fig. 6d constitute a standard stereographic triangle, which is almost within the central converged area.

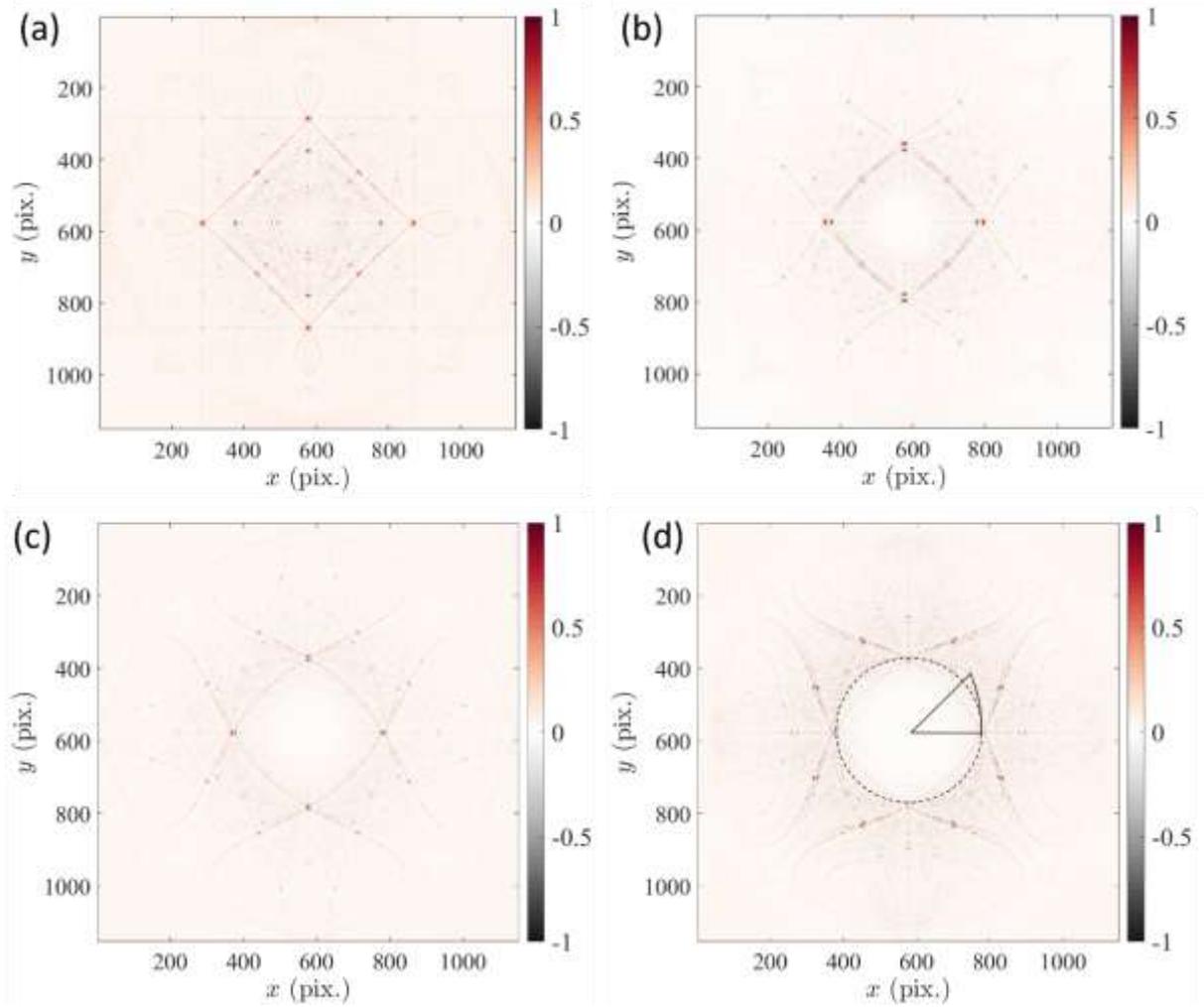

**Fig. 6.** (a)-(d) show the difference maps between the simulation results of the MS1-4 and the MS5, respectively.

It can be observed that the increased precision of the numerical solution, resulting from raising the order of the Taylor expansion, does not improve the calculation accuracy for all bands equally. Instead, it expands the radius of the trustworthy region, centered on the incidence point, where the calculated bands conform to a perfect spherical projection. Furthermore, the difference increases gradually. The outermost edge of all difference maps shows a relatively small residual, attributed to the fact that the peripheral patterns in the MS patterns inherently possess low contrast and intensity, as shown in Fig. 5.

Yang demonstrated that although the RS method is computed in real space and the BW method in reciprocal space, the difference in precision between them is minor [34]. A quantitative calculation of the exit wave functions for four different crystal structures showed that the RS method is marginally

14 / 29

superior to the BW method in terms of simulation precision. Thus, it can be concluded that the band details calculated by the low-order MS method already exhibit high precision, and the passage from the low-order to the high-order MS method primarily affects only the size of the perfect spherical projection region.

## 3.2   Radial correction of MS simulation

The kinematic simulation was the first proposed method for EBSP simulation. It successfully predicts the brightness ordering of Kikuchi bands and their positions, although the gray-level simulation is less realistic. The realistic brightness distribution of the MS pattern and the precise Kikuchi band positions from the kinematic simulation can be combined to produce a reliable master pattern. This study developed a specialized DIC tool to evaluate the radial distortion in MS patterns. The objective of this tool is to minimize the following objective function:

$$\theta(P) = \sum_{x \in \text{ROI}} \left[ f(x) - g\left(x + d(P, |x|)\frac{x}{|x|}\right) \right]^2 \tag{7}$$

where $f(x)$ is the kinematic simulation, $x = (x, y)$ is the coordinate value on the hemisphere, $g(x)$ the MS pattern, $P$ the radial distortion parameter, and $d(P, |x|)$ is the radial distortion model. Note that $d(P, |x|)\frac{x}{|x|}$ enforces an isotropic distortion of the pattern. As the central parts are well predicted and the K-band mismatch increases rapidly at the periphery (shown in Fig. 5), this study adopts a power law distortion model, namely

$$d(P, |x|) = A|x|^n \tag{8}$$

Hence, two parameters $P = [A, n]$ need to be determined. The IDIC framework proposed in Ref. [35] is adopted for the calibration of $P$.

The radial distortions for the MS1-5 patterns relative to the kinematic pattern are shown in Fig. 7, whose origin point corresponds to the center of the pattern. The curves reveal that the MS1 simulation exhibits significant radial distortion, which explains why the MS1 pattern differs significantly from those of the higher-order MS methods. Starting from MS2, each increase in order reduces overall radial distortion, and the central region with near-zero distortion progressively expands. However, the effect of this improvement diminishes with each order increment. The radial distortion of the MS5 pattern has hardly changed compared to that of MS4. The blue dashed line represents the zero distortion line, while the black dashed line marks the farthest position of the standard stereographic triangle. The radial distortion of the MS5 pattern, evaluated with different parameters $A$ and $n$, after distortion correction, is practically a horizontal line. Therefore, the distortion-corrected MS5 pattern can ensure K-band accuracy through radial distortion correction and by fully exploiting crystal symmetry.

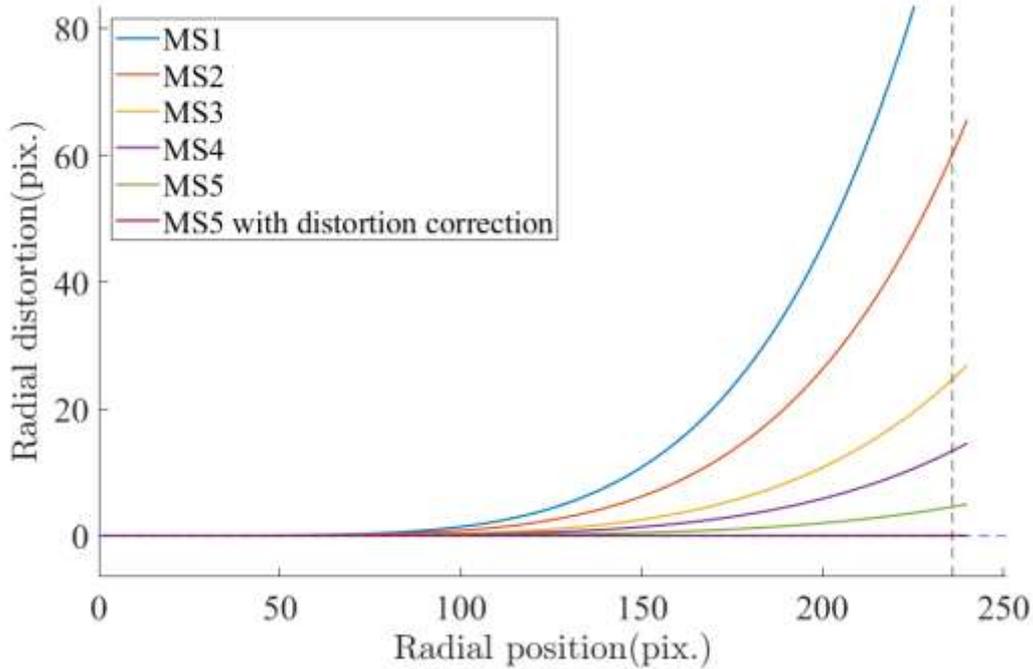

**Fig. 7.** Radial distortion curves of the MS1-5 patterns and the distortion-corrected MS5 pattern relative to the kinematic pattern. The radial position and distortion are both expressed in pixels.

We can now directly compare the kinematic simulation (Fig. 8b) with the dynamic simulations by BW (Fig. 8a) simulated using EMsoft software [36], MS5 (Fig. 8c) and the MS5 simulation with distortion correction and crystal symmetry operations (Fig. 8d, denoted 'MS5 master pattern' hereinafter). The kinematic simulation tends to overestimate contrast, whereas the MS and BW patterns exhibit more realistic Kikuchi poles and bands. The MS5 master pattern is generated from MS5 simulation by 2 steps:

I) Correct the radial distortion as expressed in Eq.8.

II) Apply the symmetry operations, which aim at generating a whole master pattern from a standard stereographic triangle located near the center (depicted in Fig. 5e). The diffraction information contained in the standard stereographic triangle is replicated 23 times and filled into the correct position by interpolation.

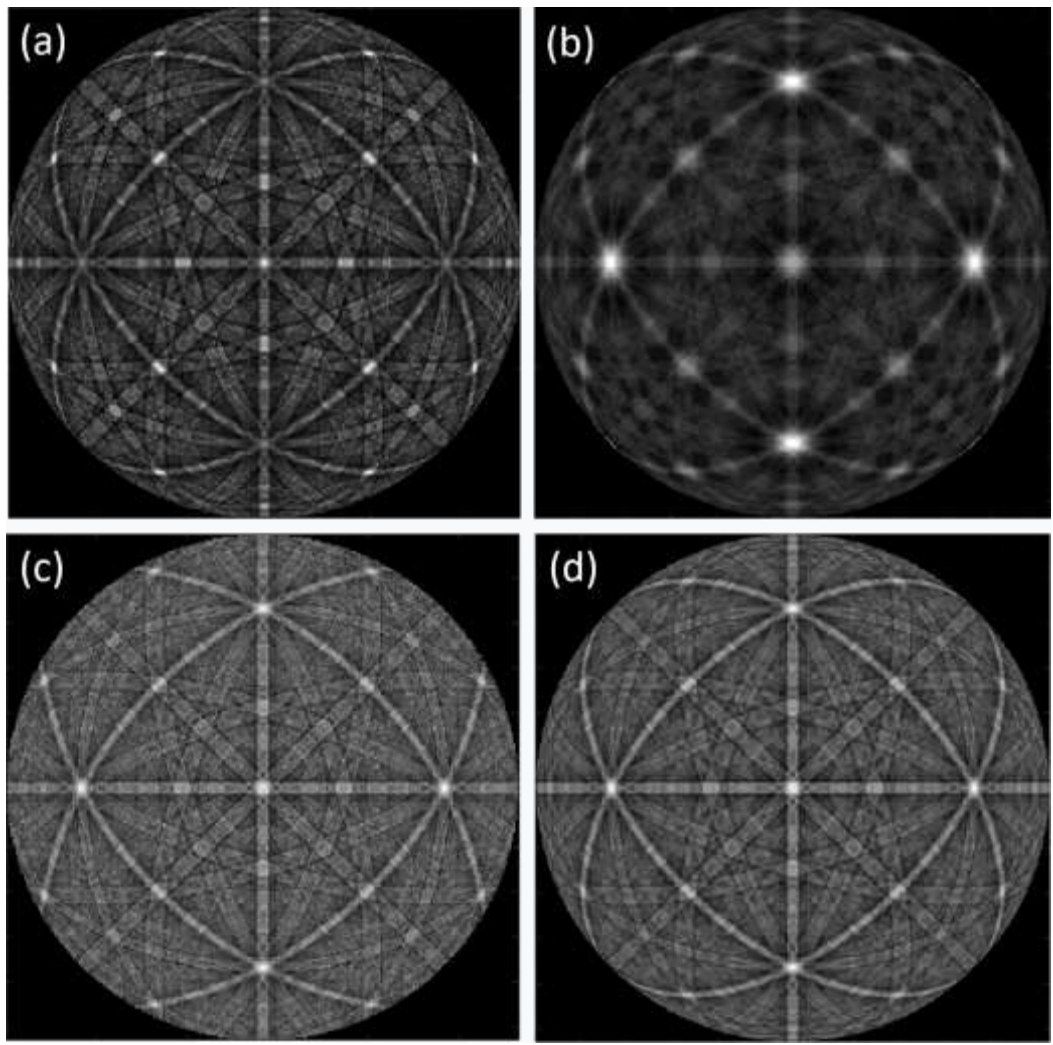

**Fig. 8.** Comparison of master patterns simulated by BW (a), the kinematic method (b), raw MS5 simulation (c), and MS5 master pattern reconstructed from standard stereographic triangle (d).

Until now, MS EBSD simulations have primarily been used for theoretical analysis of the impact of various defects [28, 29], without quantitative comparisons or indexing of experimental patterns. We now adopt the IDIC indexation technique [31] to match the experimental patterns with MS simulations at different orders and the BW master pattern. This allows us to analyze the influence of increasing the MS order on the indexing of experimental patterns and the differences between high-order MS and BW simulations during indexing.

## 3.3 Comparison with experimental pattern

Due to the low quality of the MS1 pattern, it is excluded from further comparison and analysis with the experimental pattern. In polar stereographic projection patterns, the IDIC technique identified the area that best matched the experimental pattern, resulting in the lowest residual [31]. The selected MS2-5, MS5 master pattern, and BW simulation regions were then subjected to the final planar projection. Fig. 9 compares all processed simulated patterns with the Al experimental pattern. In the MS2 pattern shown in Fig. 9b, the Kikuchi bands $(1\bar{1}\bar{1})$, $(1\bar{1}1)$, and $(11\bar{1})$, which are marked by red and blue lines in Fig. 9a, exhibit apparent curvatures. As the MS order increases to the 3rd and 4th, the $(11\bar{1})$ Kikuchi band straightens, while the (111) and (111) bands remain slightly curved. All bands, except for $(1\bar{1}\bar{1})$, are essentially straight in the MS5 simulated pattern shown in Fig. 9e. Compared to the BW simulation results shown in Fig. 9g, the difference in the position of all Kikuchi poles and bands is minimal. Fig. 9f exhibits the same K-bands and poles position as the BW simulation results, due to the radial distortion correction.

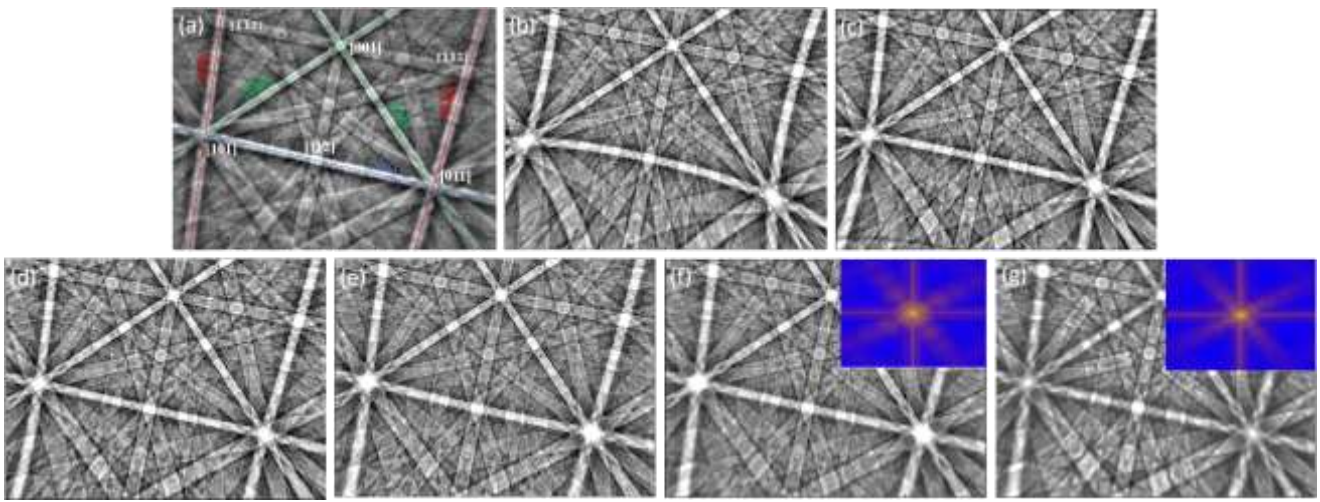

**Fig. 9.** (a) Experimental EBSP of Al-Mg with multiple Kikuchi bands indexed. (b)-(e) are the results from the MS2 to MS5 simulations, respectively, (f) is the MS5 master pattern simulation, and (g) is the BW simulation. The log-power spectrum of (f) and (g), calculated by two-dimensional (2D) fast Fourier transforms, is shown in the insets. All simulated patterns (b-f) are projected onto the crystal orientation of (a) calibrated by IDIC.

The MS patterns exhibit more precise and prominent details and band edges than the BW pattern. In the experimental pattern shown in Fig. 9a, the $(0\bar{2}0)$, $(\bar{2}00)$, and $(11\bar{1})$ bands, marked by green and blue straight lines, exhibit prominent and sharp band edges, and these three Kikuchi bands also possess higher intensity compared to the surrounding bands. However, these observations are not satisfied in the BW simulated pattern. In contrast, within the MS5 pattern, these three Kikuchi bands exhibit distinctly more prominent band edges and relatively higher intensity. It can also be observed from the two-dimensional (2D) fast Fourier transforms that Fig. 9f shows more high-frequency signals, indicating that the overall Kikuchi band edges in the MS5 simulation are sharper than those in the Bloch pattern.

The residual maps, namely the differences between the experimental pattern (shown in Fig. 9a) and the matched patterns of MS2-5, the MS5 master pattern, and BW simulations, are shown in Fig. 10. The residual norm is also calculated to quantify the pattern similarity. Table 2 lists the norm of residual values for the MS2-5 in Fig. 10 and the corresponding MS master patterns. The patterns near the [001] Kikuchi pole show little difference in the residual maps from MS2-5 methods. However, the residual of the MS2 pattern is primarily concentrated in the lower half of the EBSD pattern, which is the region far from the [001] Kikuchi pole. The band curvature outside the perfect-spherical-projection area resulted in high residuals, with an overall norm of 21.5% in Fig. 10a.

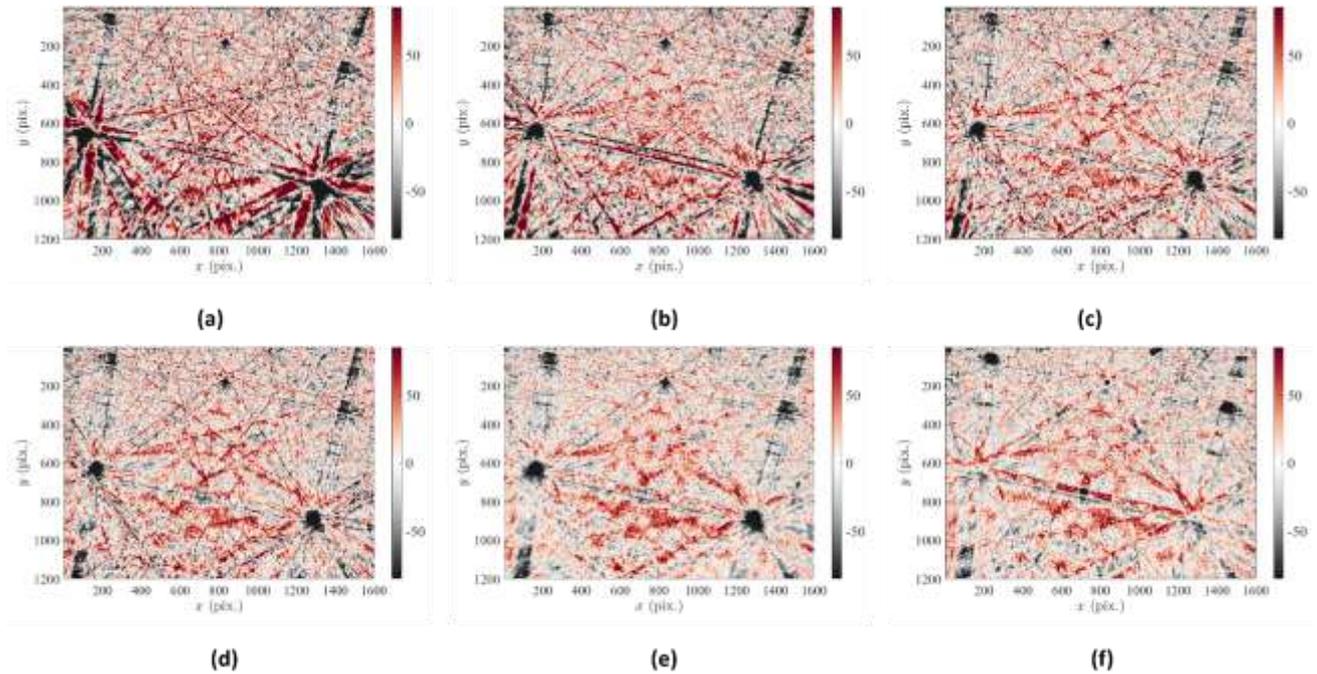

**Fig. 10**. Residual maps of the simulated patterns with the Al experimental pattern. (a) 2nd-order MS, (b) 3rd-order MS, (c) 4th-order MS, (d) MS5, (e) MS5 master pattern and (f) the BW method.

As the order of the MS method increases, the residual due to band-position discrepancies decreases, and the overall residual decreases accordingly. The residual value of the MS3 pattern was reduced by 2.7% compared to MS2, and MS4 was further reduced by 2.5%, while MS5 was merely reduced by 0.6%. This demonstrates that the positions of the Kikuchi poles and bands in the MS5 method simulated pattern do not change significantly compared to MS4, indicating that the benefits of the higher-order Taylor expansion (Equation 2) tend to saturate at MS5. Given that calculating the next Taylor expansion order triples the computing time, the MS maximum order is set to 5.

Table 2: The norm of residual values for the 2nd to 5th-order of MS raw patterns and master patterns. The residual of the Bloch wave simulation is provided for comparison.

|  | MS2 | MS3 | MS4 | MS5 | Bloch wave |
|---|---|---|---|---|---|
| Raw pattern | 21.5% | 18.8% | 16.3% | 15.7% | - |
| Master pattern | 16.5% | 15.1% | 14.3% | 13.7% | 12.9% |

Calculating with MS master patterns can further reduce the residuals compared to MS raw patterns. Lower-order MS methods show a greater reduction in residuals, as the erroneous peripheral bands are replaced by a relatively better simulated standard stereographic triangle. However, the residual value of the MS5 master pattern remains at 13.7%, higher than the 12.9% in the BW simulation. A comparison of their residual maps (Figures 10e and f) reveals that the higher residual value for MS5 is actually due to intensity differences across the [001], [101], and [011] Kikuchi poles. Conversely, for the [112], [1$\bar{1}$2], and [$\bar{1}$12] Kikuchi poles, the MS simulation exhibits lower residual values.

## 4. Application of MS5 simulation to experimental pattern indexation

The MS5 master pattern is used as a reference for IDIC to index experimental EBSPs. The IDIC algorithm is based on the assumption that an experimental pattern is projected based on only six geometric parameters: three coordinates of the projection center $(x^*, y^*, z^*)$, and Euler angle triplet $(\varphi_1, \phi, \varphi_2)$ (for detailed explanations, refer to [31]). Two EBSD datasets of different qualities are analyzed and compared in the following.

### 4.1 Indexation of high-quality Al alloy experimental patterns

High-resolution EBSPs of 1200×1600 pixels of an annealed Al-Mg alloy, recorded with a Bruker e-FlashHD detector and reported in [31], were used for validation (download link: https://doi.org/10.5281/zenodo.6990325). The EBSPs were indexed by matching them with the raw

MS5 simulation (without distortion correction) and the MS5 master pattern. The difference between Al and Al-Mg alloy has been checked by MS simulation not to involve any quantitatively appreciable change in the diffraction pattern. Their indexation results were compared with those of the well-established BW-simulation full pattern match [37], as shown in Fig. 11a. The misorientation is segregated based on grain orientations, with a maximum value of 3° for the uncorrected MS5 simulation with few pixels, the great majority of pixels are within 1.1°, an error level comparable to that of the conventional EBSD technique [38]. The misorientation is correlated with crystal orientation, as shown in Fig. 11b, because the distortion increases as the reference pattern selected moves further from the [001] Kikuchi pole. The MS5 master pattern indexation results in misorientations of all values below 0.2°, with an average value of 0.102°. It should be noted that by combining the brightness simulation of MS5 with the kinematically predicted Kikuchi band positions, satisfactory master patterns can be formulated, resulting in EBSD indexation accuracy comparable to that of the state-of-the-art BW method. Therefore, experimental patterns can be precisely indexed using the MS5 master pattern.

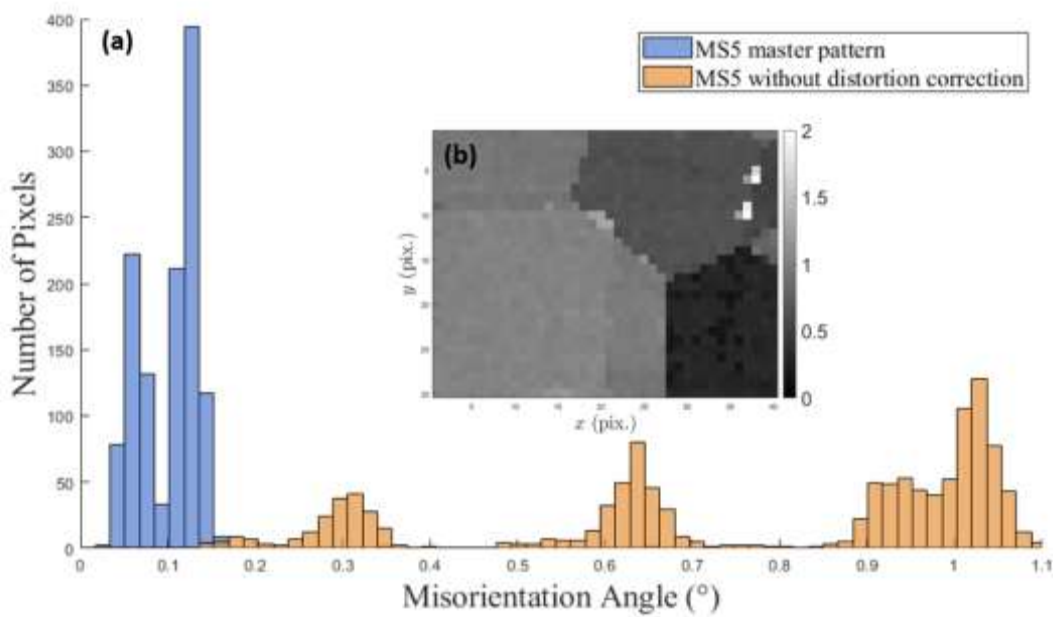

**Fig. 11.** (a) Histogram of the misorientation against the BW-based indexation of the annealed Al-Mg alloy patterns, calculated via IDIC using the MS5 master pattern and raw MS5 simulation (without distortion correction). (b) Misorientation map between indexations with raw MS5 and BW simulations.

The KAM map frequently highlights the sample local misorientation and benchmarks different indexation methods. The KAM for these experimental patterns, using the IDIC indexation with master patterns of BW and MS5, was calculated. To observe angular differences within the grains more clearly, all orientation differences greater than 0.8° or 2° have been removed, as shown in Fig. 12.

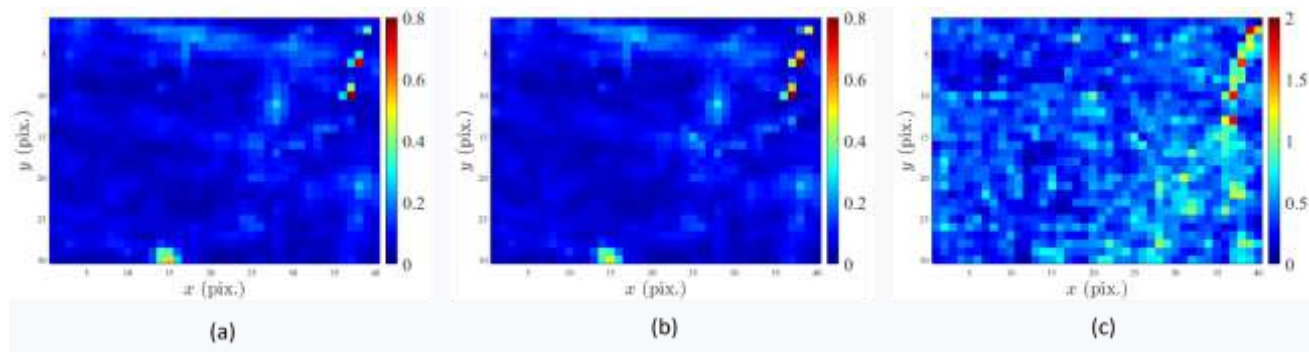

**Fig. 12.** KAM maps of the Al-Mg alloy experimental patterns calculated using (a) the BW master pattern, (b) the MS5 master pattern, and (c) a Hough-transform-based commercial software.

Comparing the KAM images calculated by the BW and MS5 patterns, we find that the KAM values for most pixels are essentially consistent. Only a few pixels differ slightly. Additionally, the KAM calculated using the commercial software shown in Fig. 12c is significantly higher than that obtained by the BW and MS5 methods. The KAM distributions for all pixels, indexed by the three methods, are plotted as a histogram in Fig. 13 to facilitate comparison. Fig. 13 shows that the results from the commercial software, based on the Hough Transform, deviate significantly from those of the other two methods. Hough-transform-based KAM values approximate a normal distribution centered around 0.4°. The overall KAM values calculated by IDIC using the master patterns MS5 and BW are both significantly lower than those of the widely used Hough Transform method. Thanks to the IDIC calibration scheme, the crystal orientation of cubic crystals can be reliably indexed by matching to the MS5 master pattern.

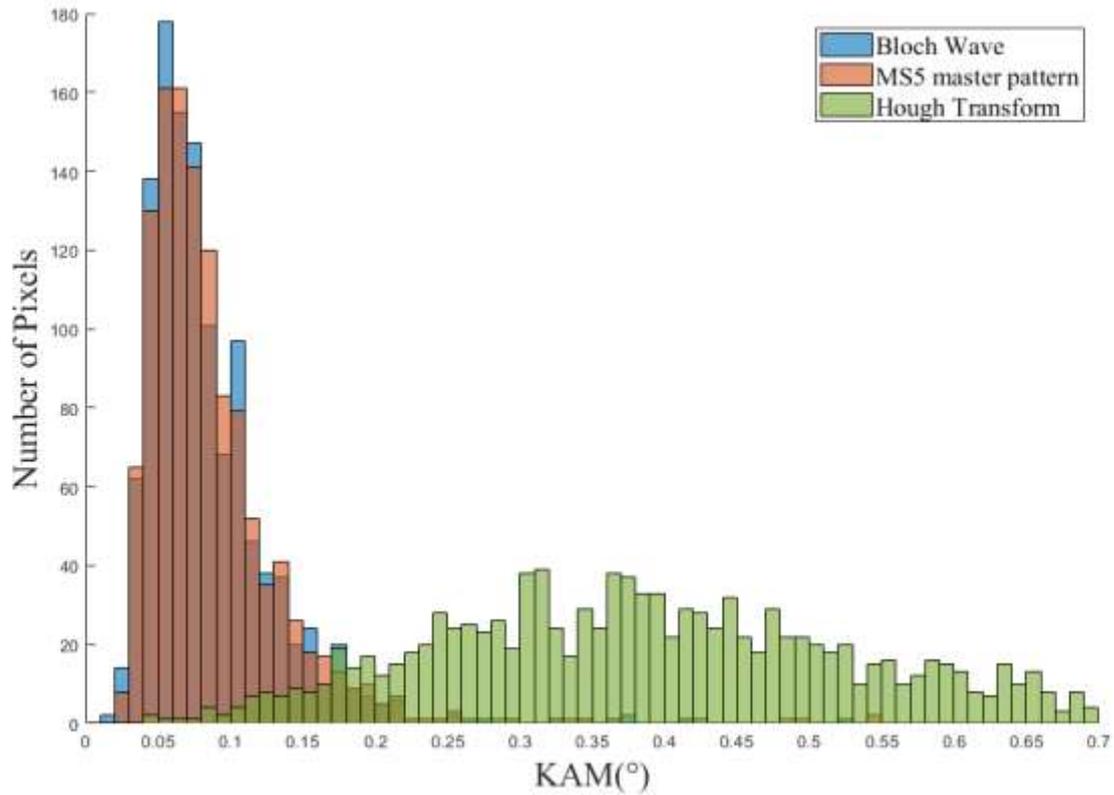

**Fig. 13.** Histogram of the KAM values from Fig. 12, indexed using the whole pattern match method with the BW master pattern, the MS5 master pattern, and the commercial software based on the Hough Transform.

## 4.2 Indexation of median-quality Al composite experimental patterns

Fast-acquisition EBSPs of Al composites are also analyzed to test the indexation performances on more realistic data reported in Ref. [39]. The detector is a Symmetry S2 EBSD, the dwell time is 0.1s, and the pattern size is 1024×1244.

For EBSD data analysis, the inverse pole figure (IPF) is a commonly used and intuitive tool for visualizing orientations. The IPF-Z of an Al-based alloy, indexed by IDIC using MS5 master pattern, Bloch master pattern, and Hough transform software, is shown in Fig. 14(a-c), respectively. The Hough transform often fails to index patterns at grain boundaries (Fig. 14c), even within several grains, which seriously impacts the determination of grain boundaries and orientations. This phenomenon arises because EBSPs at grain boundaries are overlapping patterns from different orientations, as shown in Fig. 14d. The Hough transform has limitations for identifying superimposed band systems of poor quality. However, when using IDIC with a simulated master pattern as a template, indexation of overlapped EBSPs remains successful by initiating with neighboring indexed pixels [4, 41]. Fig. 14e shows the simulated pattern corresponding to one set of Kikuchi bands of Fig. 14d. The high similarity of Fig. 14a and Fig. 14b demonstrates that the MS5 master pattern is sufficient for indexing experimental patterns, comparable to the well-established BW master pattern.

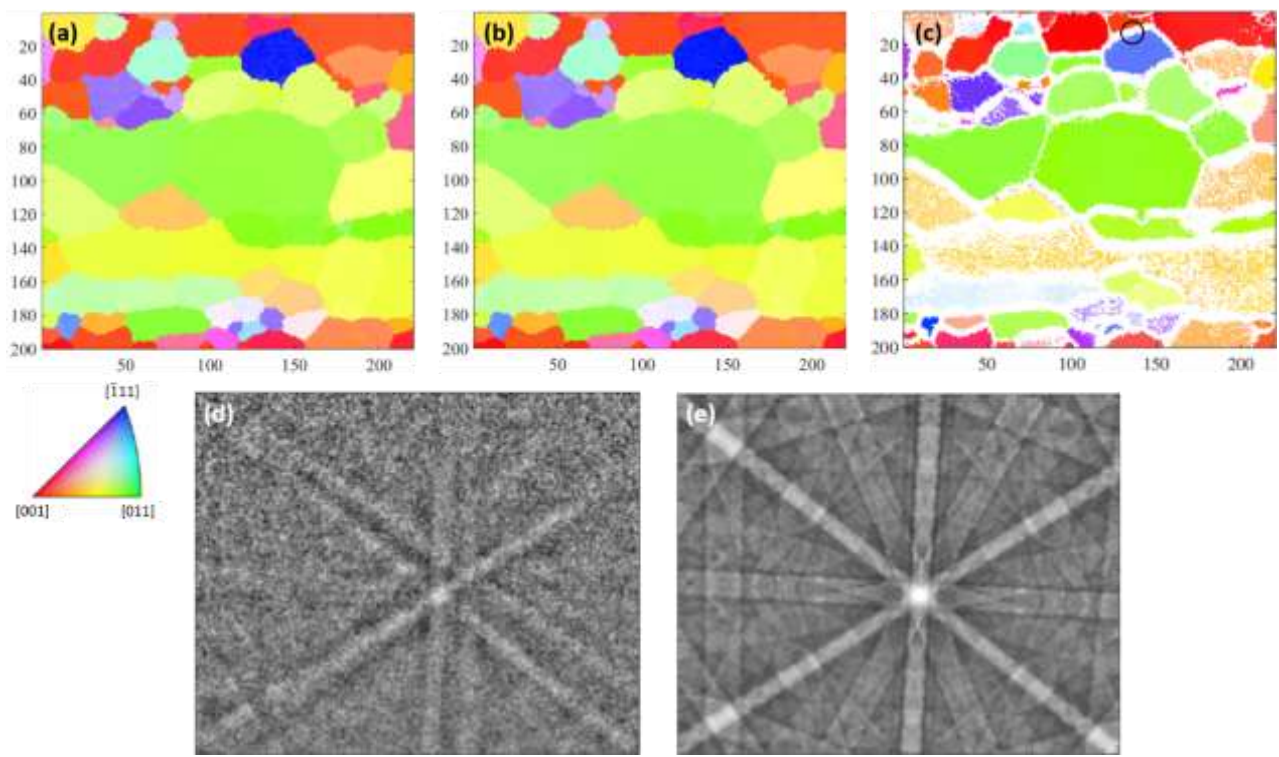

Fig. 14. Inverse pole figure of the Al-based alloy experimental patterns calculated using (a) the MS5 master pattern, (b) the BW master pattern, and (c) a Hough-transform-based commercial software. (d) The experimental pattern at the grain boundary marked by a circle in Fig. 14c and (e) the corresponding projected pattern from the MS5 master pattern.

## 5. Conclusion

The MultiSlice (MS) method is a crucial branch of electron diffraction simulations, offering a unique advantage in computing the diffraction patterns of crystal models with defects. Although EBSD simulation by MS has undergone continuous development, it has not been quantitatively compared with experimental patterns, let alone used for crystal orientation determination.

In this work, the high-energy assumption was not applied, as the acceleration voltage was lower than 30kV, which is standard practice for EBSD. The generic expression (for any order) of Taylor's expansion in the MS framework is derived to numerically approach the square root operation necessary in the stationary Schrodinger equation. Vectorization and GPU acceleration in Matlab speed up the MS simulation. It is found that increasing the expansion order leads to a diffraction pattern with a gradually expanding reliable central area, as compared with the well-established Bloch Wave (BW) simulation results. The 5th-order implementation of MS, MS5, is chosen as it compromises computation time and pattern precision. The outer regions of the MS simulation can be corrected by a specifically designed radial distortion correction technique. Furthermore, the crystal symmetry is fully employed to replicate the inner region onto the peripheral zones to obtain an MS5 master pattern. The MS5 master pattern is subsequently projected onto the detector screen, enabling direct comparison with experimental patterns for the first time.

Using both the BW and MS5 master patterns, an integrated digital image correlation (IDIC) technique was adopted to calculate and compare the misorientation of Al-Mg experimental patterns across various orientations. The misorientation between the results with BW and MS5 master patterns is well within 0.2°. Additionally, the calculated KAM maps and IPFs for the experimental patterns, obtained using both simulation master patterns, exhibited undiscernible discrepancies. This demonstrates that the MS5 master pattern is sufficient to index experimental patterns of cubic crystal structures with high precision. The MS EBSD simulations, now optimized to be comparable with experimental patterns, offer an accuracy nearly identical to that of the BW method. The only limitation is its high computational cost, which is expected to decrease as computational capability continues to grow on modern computers. The unique advantage of MS in simulating imperfect crystals is expected to open new perspectives for analyzing crystal defects, such as the distribution and density of dislocations and nanotwins.

## Acknowledgement

This work is financially supported by the National Natural Science Foundation of China [No. 52273229, 12074326, 51901132] and the MAI (Materials Ageing Institute, http://themai.org).

## Data Availability

The electron diffraction patterns of the annealed Al-Mg sample analyzed in the paper are available at Zenodo (https://doi.org/10.5281/zenodo.6990325).